\documentstyle[preprint,eqsecnum,aps]{revtex}
\begin{document}
\draft
\preprint{\vbox{\baselineskip=12pt
\rightline{CGPG-93/8-4}
\rightline{hep-th/9308141}}}
\title{Poisson Brackets on the Space of Histories}
\author{Donald Marolf\cite{Marolf}}
\address{Physics Department, The Pennsylvania State University,
University Park, PA 16802} \date{revised February, 1994}
\maketitle

\begin{abstract}

We extend the Poisson bracket
from a Lie bracket of phase space functions
to a Lie bracket of functions on the space of canonical
histories and
investigate the resulting algebras.  Typically, such
extensions define corresponding Lie
algebras on the space of Lagrangian histories via pull back
to a space of partial solutions.  These are the same spaces
of histories studied with regard to path integration and
decoherence.  Such spaces of histories are familiar from
path integration and some studies of decoherence.  For gauge  systems,
we extend both the canonical and reduced Poisson brackets
to the full space of histories.  We then comment on the
use of such algebras in  time reparameterization
invariant systems and systems with a Gribov ambiguity,
though our main goal is to introduce concepts and
techniques for use in a companion paper.
\end{abstract} \pacs{}

\section{Introduction}
\label{intro}

Formulations of quantum theories can be roughly divided
into two classes.  The first follows the algebraic
approach in which a commutator  *-Algebra is defined and a
Hilbert space representation is sought, while the second
defines ``transition amplitudes" by path integration over
some space ${\cal H}$ of histories and then interprets
these amplitudes as matrix elements of operators in a
Hilbert space.  Both here and in the companion paper
\cite{other}, we will be concerned with the algebraic
approach and, more specifically, with the classical
(commuting) *-Lie algebra on which the quantum commutator
is often based.

This classical algebra is the algebra of complex functions
on some space with the usual operations of multiplication,
addition, and complex conjugation  (*) supplemented by a
Lie bracket operation.  For gauge-free systems,  this may
be the Poisson algebra ${\cal A}_H(\Gamma)$ of complex
functions on the phase space $\Gamma$ or the Peierls
algebra ${\cal A}_L({\cal S})$ of functions on the space
${\cal S}$ of solutions to the equations of motion.  The
subscripts $H$ and $L$ refer to the Hamiltonian and
Lagrangian methods associated with the construction of
these algebras. ${\cal A}_H(\Gamma)$ and ${\cal A}_L({\cal
S})$ are isomorphic \cite{Peierls} under any map that
takes ${\cal S}$ to $\Gamma$ by evaluating the phase space
coordinates at some time $t$.

For systems with gauge symmetries we may pursue either
option in the full theory and, in addition, we may choose to
keep the gauge symmetry
intact, or we may take the quotient by gauge
transformations and consider reduced dynamics. This leads
to two Poisson algebras, ${\cal A}_H(\Gamma)$ of Dirac
\cite{Dirac} on the full phase space $\Gamma$ and ${\cal
A}_H(\Gamma_r)$ on the reduced phase space $\Gamma_r$, as
well as two Peierls algebras, ${\cal A}_L({\cal S}_r)$ on
the space ${\cal S}_r$ of reduced solutions and ${\cal
A}^{GI}_L({\cal S})$ defined on gauge invariant functions
on ${\cal S}$. The Peierls bracket of gauge dependent
functions is not defined.

Each algebra has advantages and disadvantages.  For
example,  ${\cal A}_H(\Gamma)$, ${\cal A}_H(\Gamma_r)$,
and ${\cal A}_L({\cal S}_r)$ include all smooth functions
on the relevant spaces and may often be expressed as the
set of suitable combinations of a small number of
functions, such as the canonical coordinates and momenta,
that have simple algebraic properties.  This greatly
simplifies factor ordering during quantization as well as
many Lie bracket computations.  However, the reduced
spaces, $\Gamma_r$ and ${\cal S}_r$, and algebras, ${\cal
A}_H(\Gamma_r)$ and ${\cal A}_H(\Gamma)$,  may be
difficult to construct, especially when gauge fixing is
impossible. Additionally, it may be desirable to display
the gauge symmetries explicitly.

For relativistic field theories, another important
distinction is that  ${\cal A}_L({\cal S}_r)$ and ${\cal
A}_L({\cal S})$ are manifestly covariant while ${\cal
A}_H({\Gamma_r})$ and ${\cal A}_H(\Gamma)$ are defined
only after a 3+1 decomposition of spacetime.  Manifest
covariance simplifies verification of the quantum theory
and facilitates such tasks as the  derivation \cite{Bryce}
of the measure for the covariant path integral. A related
remark is that, since ${\cal A}_H(\Gamma)$ and ${\cal
A}_H(\Gamma_r)$ are defined on functions at some single
time, they lead most naturally to Schr\"odinger picture
quantum mechanics while the Peierls bracket leads most
directly to Heisenberg picture quantization since ${\cal
A}_L({\cal S}_r)$ and ${\cal A}_L({\cal S})$ include
functions defined at all times and even functions with
nonlocal spacetime support.  We note that a Heisenberg
picture is particularly useful for a time
reparameterization invariant system since it allows
reparameterization invariant operators to be defined by
integration over time.

Despite these differences,
three of these algebras are isomorphic.  Functions on
${\cal S}_r$ are just gauge invariant functions on ${\cal
S}$ and  evaluating the reduced phase space coordinates at
some time $t$ maps ${\cal S}_r$ to ${\Gamma}_r$.  ${\cal
A}_H(\Gamma)$ is different, however, as some points in
$\Gamma$ do not correspond to solutions or points in
$\Gamma_r$ and evolution on $\Gamma$ is not uniquely
defined.   We will find it useful to partially bridge this
gap by defining  a space ${\cal E}$
of all possible evolutions of points in $\Gamma$.

Algebraic structures on these spaces have been compared
in \cite{Wald}, \cite{Abhay}, and \cite{Witten}. Of
particular interest is \cite{Wald} which considers
structures on the space ${\cal H}$ of histories as well.
${\cal H}$ is the domain space of the system's action
functional so that it is typically the space of suitable
fields on some manifold $M$.  This space is called ${\cal
F}$ in \cite{Wald} and $\Phi$ in \cite{Bryce}; we use
${\cal H}$ to avoid confusion with other notation
introduced here and in \cite{other}. ${\cal H}$ is an
especially useful space to consider since it contains
both ${\cal S}$ and ${\cal E}$ as subspaces and from these
${\Gamma}$, ${\Gamma}_r$, and ${\cal S}_r$ can be reached
by projection.  In addition, studies of structures on ${\cal H}$ may
help to connect the algebraic and path integral approaches
to quantization.

References \cite{Wald}, \cite{Abhay}, and \cite{Witten}
discuss a presymplectic structure on ${\cal S}$ and ${\cal
H}$.  While useful for other purposes, the presymplectic
form is degenerate and cannot be inverted to define a Lie
bracket of functions on ${\cal S}$ and ${\cal H}$.
Quantization on ${\cal S}$ and ${\cal H}$ cannot proceed
until this degeneracy is removed.

Our goal here and in \cite{other} is the construction of
Lie algebras  ${\cal A}_L({\cal H})$ of
complex functions on ${\cal H}$ and thus on ${\cal E}$ and ${\cal S}$
by pull back and on $\Gamma$, $\Gamma_r$, and ${\cal S}_r$ by projection.
As with the presymplectic structure, these
algebras reflect the gauge structure -- often by becoming
degenerate.  As a result, the contravariant tensor field
that defines the Lie bracket cannot be inverted to define
a symplectic form.  Thus, our study is complementary to
that of \cite{Wald}, \cite{Abhay}, and \cite{Witten}.

The construction of these algebras will be presented in
\cite{other} and is based on the Peierls bracket.  In
particular, it extends ${\cal A}^{GI}_L({\cal S})$ to
${\cal A}^{GI}_L({\cal H})$ and then generalizes this
algebra to ${\cal A}_L({\cal H})$ using the methods of
\cite{Bryce}.  However, much of this development can be
described in terms of extensions ${\cal A}_H({\cal H})$ of
the Poisson algebras ${\cal A}_H(\Gamma)$ and ${\cal
A}_H(\Gamma_r)$ to ${\cal H}$.   Such a description serves
two purposes.  By acting as an intermediate step, it
greatly facilitates comparisons of ${\cal A}_L({\cal H})$
with ${\cal A}_H(\Gamma)$ and ${\cal A}_H(\Gamma_r)$ and
thus with conventional quantization methods and, because
the Poisson bracket is more familiar than the Peierls
bracket, it allows the introduction of useful concepts,
such as the space ${\cal E}$ of evolutions, and
techniques, such as defining algebras locally, pulling
back algebras from ${\cal H}$, and ``gauge breaking"
without heavy machinery.  This is the subject of the work
below.

Our study begins in section \ref{uncon} where ${\cal
A}_H({\cal H})$ is introduced for gauge-free systems.  For
gauge systems, ${\cal A}_H({\cal H})$ is defined in
\ref{constraints} as an extension of ${\cal A}_H(\Gamma)$
and in \ref{gauge fixing} as an extension of ${\cal
A}_H(\Gamma_r)$ using ``gauge breaking" -- a sort of
generalization of gauge fixing that may be performed in
the presence of a Gribov ambiguity.  Because the Poisson
bracket is grounded in the canonical formalism, sections
\ref{uncon} and  \ref{Gauge} refer to the space ${\cal H}$
of canonical histories of phase space coordinates, but
section \ref{Lagrange} shows that such extensions
also exist on spaces of so-called Lagrangian
histories defined by more general fields.  Section
\ref{Quant} explores the implications of using ${\cal
A}_H({\cal E})$ and ${\cal A}_H({\cal S})$ for
quantization, comparing this with quantization of  ${\cal
A}_H(\Gamma)$ and ${\cal A}_H(\Gamma_r)$.  We then close
with  a summary discussion and two appendices.  Appendix A
constructs a globally defined Lie algebra of functions on
a manifold from a set of compatible algebras defined
locally and appendix B describes the extension of the
Dirac bracket to ${\cal H}$.

\section{Unconstrained Systems}
\label{uncon}
In this section we describe the extension of the Poisson
bracket to  a Lie bracket on the
space ${\cal H}$ of histories for a system presented in
unconstrained canonical form.  Recall that a Lie bracket
is a bilinear  antisymmetric operation $(,)$ that
satisfies the Jacobi identity and the derivation
requirement: \begin{equation}
\label{derivation req}
(AB,C) = A(B,C) + (A,C)B
\end{equation}
Consider a system with no gauge
symmetries which is kinematically described by a
phase space $\Gamma$ during the
time interval $I = [t_i,t_f]$.  Histories of this system
lie in the space $\Gamma^I$ given by the direct product
of one copy $\Gamma_t$ of the phase
space $\Gamma$ for each time $t$ in  the interval $I$.
If $z^A$ for $A$ in some index set ${\cal I}$ are
canonical coordinates on $\Gamma$ then
$z^i$ for $i = (A, t) \in {\cal I} \times I$ are
coordinates on $\Gamma^I$.
We take the dynamics of this system to be governed by a
Hamiltonian $H(t) = H(z^A(t),t)$ and a Poisson bracket
$\{z^A,z^B\} = \Omega^{AB}$ or, equivalently, by a first
order action principle of the form \begin{equation}
\label{canact}
S(z^i) = \int_{t_i}^{t_f} dt (\case{1}{2}
\Omega^{-1}_{AB} z^A(t) \dot{z}^{B}(t) - H(t))
\end{equation}
for some time and field independent invertible
antisymmetric matrix $\Omega^{AB}$.  We now define our
space ${\cal H}$ of histories to be that subspace of
$\Gamma^I$ that lies in the domain of ${\cal S}$.

As described in the introduction, the phase space $\Gamma$
is isomorphic  to the space ${\cal S}$ of solutions.  This
isomorphism carries the Poisson bracket $\{,\}$ to a Lie
bracket $(,)_{\cal S}$ on ${\cal S}$. We seek an extension
of the Poisson bracket to ${\cal H}$ in the sense that any
Lie bracket $(,)_{\cal H}$ that we define on ${\cal H}$
should have a well-defined pull back through the inclusion
map $i: {\cal S} \rightarrow {\cal H}$ that coincides with
$(,)_{\cal S}$:

\begin{equation}
\label{pull back}
\{F \circ i, G \circ i \}_{\cal S} = (F,G)_{\cal H} \circ
i \end{equation}

If this is to be
well-defined, we must have $(F,G)_{\cal H} \circ i =
(J,K)_{\cal H}  \circ i$ whenever
$F \circ i = J \circ i$ and $G \circ i = K \circ i$.
Equivalently, we could have $(F,G)_{\cal H} \circ i = 0$
whenever $F \circ i = 0$ and in particular $(S,_i,A)_{\cal
H}  \circ i = 0$ for all $A$.  Here, the comma denotes the
(functional) derivative with respect to the coordinate
$z^i$ and we use the condensed notation of \cite{Bryce}.

Assuming that $S$ is sufficiently smooth and $S,_i$ is
appropriately behaved, the condition $(S,_i,A)_{\cal H} =
0$ would be  enough to guarantee that
the bracket of sufficiently smooth functions pulls back to
${\cal S}$. However, we choose to impose the stronger
condition that $(S,_i,A)$ should vanish identically on
${\cal H}$. We will refer to this as the
condition for $(,)_{\cal H}$
to ``respect" the equations of motion $\{S,_i = 0\}$.
Note that this choice is coordinate dependent as it is not preserved
under passing to new coordinates $Z^j = Z^j(z^i)$ unless
${{\partial z^i} \over {\partial Z^j}}$ has identically
vanishing bracket. If this change of variables ultralocal
in time,  \ref{pull back} implies that the transformation
is linear.

This coordinate dependence could be removed by choosing a
suitable set of functions $\{f\}$ on ${\cal H}$ that
vanish on ${\cal S}$ to satisfy $(f,A)_{\cal H} = 0$ for
all $A$ in place of $\{S,_i\}$.  The coordinate
dependence thus becomes a dependence on the set of
functions chosen.  However, because the definition of
``suitable" functions is complicated and  because the form
of the equations of motion $S,_i$ defined by the
coordinates $z^i$ is particularly convenient, we will  use
the coordinate dependent formulation. This also eases
comparison with \cite{other} in which coordinate
independent extensions of the Peierls bracket are defined
by torsion-free connections instead of by sets of
functions.

Appendix A shows how a Lie bracket of functions on a
manifold $M$ can be assembled from a set of Lie brackets
of functions on patches $U_i  \subset M$ when these
algebras agree in the overlap regions $U_i \cap U_j$.  To
ensure this agreement,  we choose coordinates defined
globally on $\Gamma$ or defined  in patches such that the
transitions functions are linear.  We will refer
to a manifold described
by such coordinates as a {\it linearized
manifold} or as a manifold with a linearized
structure.  While it is not entirely
clear which differential manifolds
are diffeomorphic to linearized manifolds, we note that
linearized structures for the sphere, torus, and other
simple but nontrivial manifolds can be readily
constructed and proceed with our discussion.
The conditions
$(S,_i,A) = 0$ are then imposed on the local algebra of
each patch.

Because of the derivation requirement \ref{derivation
req}, which implies that  $(F,G)_{\cal H} = F,_i
(z^i,z^j)_{\cal H}G,_j$,  the extended bracket is entirely
determined by its action $(z^i,z^j)_{\cal H}$ on
coordinate functions.   Since $(z^A(t),z^B(t))_{\cal H}
\circ i  = \{z^A(t),z^B(t)\}_{\cal S} \equiv \Omega^{AB}$
and $\Omega^{AB}$  is a matrix of
constant functions on ${\cal S}$, we define $(z^A(t),z^B(t))_{\cal H}$ to
be the corresponding matrix of constant functions on ${\cal H}$.  The
use of this extension represents another choice that we have
made, but this time a coordinate independent one.

{}From this choice, the bracket $(z^i,z^j)_{\cal H}$ of
coordinate functions at different times is specified
uniquely by our condition that the algebra respect the
equations of motion $\{S,_i = 0\}$. To see this, we note
that for $i = (A,t)$ we have: \begin{equation}
\label{bracketprop1}
0=(S,_i,z^B(t'))_{\cal H} = (\Omega^{-1}_{CA}\dot{z}^A(t)
- H_{|C}(t), z^B(t')) _{\cal H}
\end{equation}
where $_{|C}$ denotes a derivative with respect to $z^C$
on $\Gamma$. It follows that
\begin{equation}
\label{bracketprop2}
0 = \case{\partial}{\partial t}(z^A(t),z^B(t'))_{\cal H}
-  \Omega^{AC} H_{|CD}(z^D(t),z^B(t'))_{\cal H}
\end{equation}
in which the combination $(\Omega\circ H)_C^B =\Omega^{BD}
H_{|DC}$ acts like a connection and propagates
$(z^A(t),z^B(t'))_{\cal H}$  from one time to another.
The solution to \ref{bracketprop2} is thus

\begin{equation}
\label{Omega mult}
(z^A(t_1),z^B(t_2))_{\cal H} = T_L(t_2,t_1)^A_C
\Omega^{CB} = \Omega^{AC} T_R(t_2,t_1)^B_C
\end{equation}
where
\begin{equation}
T_L(t_2,t_1)^A_C = {\cal P} \exp [\int_{t_2}^{t_1}dt \
\Omega\circ H]^A_C, \qquad T_R(t_2,t_1)^B_C = {\cal P}
\exp [\int_{t_2}^{t_1}dt \ H \circ \Omega]^B_C
\end{equation} and ${\cal P}$ denotes path ordering.  Our
bracket $(,)_{\cal H}$ is then {\it defined} by \ref{Omega
mult} and \ref{derivation req}.

To see that $(,)_{\cal H}$ is in fact a Lie bracket on
${\cal H}$, note that the derivation property is manifest
from the construction and that antisymmetry and the Jacobi
identity will follow if we establish that $(z^i,z^j)_{\cal
H} = - (z^j,z^i)_{\cal H}$ and  \begin{equation}
\label{cyclic}
\sum_{i,j,k \in {\cal T}}
\epsilon_{ijk}
((z^i,z^j)_{\cal H},z^k)_{\cal H} = 0
\end{equation}
for any
three element subset ${\cal T}$ of ${\cal I} \times I$
and any antisymmetric symbol $\epsilon_{ijk}$ on ${\cal T}$.
The  symmetry of $H_{|AB}$ implies that $H_{|AC}\Omega^{CB}
= - \Omega^{BC} H_{|CA}$ and we have the useful property:
\begin{equation}
\label{transpose}
{\cal P} \exp [\int_{t_1}^{t_2}dt \ H\circ \Omega]^B_C =
{\cal P} \exp [\int_{t_2}^{t_1}dt  \ \Omega \circ H]^B_C
\end{equation}
Antisymmetry then follows directly:
\begin{equation}
\label{Hext}
(z^A(t_1),z^B(t_2))_{\cal H} = \Omega^{AC}
T_R(t_1,t_2)^B_C = - T_L(t_2,t_1)^B_C \Omega^{CA} = -
(z^B(t_2),z^A(t_1))_{\cal H}   \end{equation}

Properties \ref{Omega mult} and \ref{transpose}
can also be used to verify
the Jacobi identity.  A short calculation shows that
\begin{equation}
((z^{A_1}(t_1),z^{A_2}(t_2))_{\cal H},z^{A_3}(t_3))_{\cal
H}  = - \int_{t_1}^{t_2} dt
\ H_{|B_1B_2B_3}(t) \prod_{i=1}^3
(z^{B_i}(t),z^{A_i}(t_i))_{\cal H} \end{equation}
so that the cyclic sum in \ref{cyclic}
vanishes due to the symmetry of
$H_{|B_1B_2B_3}$.
We have thus succeeded in extending the Poisson bracket
for an unconstrained system through \ref{Omega mult} and
\ref{derivation req} to a Lie bracket
$(,)_{\cal H}$ on ${\cal H}$.  We turn now to systems with
gauge symmetries and constraints.

\section{Gauge Systems}
\label{Gauge}
This section develops extended Poisson algebras for gauge
systems and other constrained systems.
Two types of Poisson algebra will be of interest,
those defined through canonical procedures \cite{Dirac}
and those defined through gauge fixing or other reduced
phase space  procedures.   Each of these will
be addressed in a separate subsection, though we will see
that the two are quite similar and that both follow from a
more general ``gauge breaking" scheme.  The last
subsection compares the canonical case with canonical
gauge fixing.

\subsection{Phase Spaces with Constraints}
\label{constraints}

Consider a system described not by an
action of the form \ref{canact}, but by an action of the
form: \begin{equation}
\label{canact gauge}
S = \int_{t_i}^{t_f} dt[\case{1}{2}
\Omega^{-1}_{AB}z^A\dot{z}^B -H_0(t) -
\lambda^a(t) \phi_a(t)]
\end{equation}
for $a$ in some index set ${\cal G}$, $A \in {\cal I}$,
$\Omega^{AB}$ an invertible antisymmetric time and field
independent matrix, $H_0(t) = H_0(z^A(t),t)$, and $\phi^a
= \phi^a(z^A(t),t)$.   For convenience, we introduce an
index $\alpha = (a,t) \in {\cal G} \times {I}$ such that
$\lambda^{\alpha} = \lambda^a(t)$.

Let the lagrange multipliers
$\lambda^a(t)$ at time $t$ take values in some space
$\Lambda$. $S$ is then a function on a space
${\cal H} \subset \Gamma^I \times \Lambda^{I}$ of paths
through $\Lambda$ and paths through $\Gamma$.
The variation of this action with
respect to the Lagrange multipliers $\lambda^{\alpha}$
enforces the constraints $\phi_{\alpha} = 0$.

It will be convenient to
regard ${\cal H}$ as a union of the subspaces
${\cal H}_{c^{\alpha}} = \{(x,c^{\alpha})|x\in \Gamma^I,
(x,c^{\alpha}) \in {\cal H} \}$ on which
$\lambda^{\alpha} = c^{\alpha}$
and to consider the spaces of solutions ${\cal
S}_{c^{\alpha}} \subset {\cal H}_{c^{\alpha}}$ on which
the restriction $S_{c^{\alpha}}$ of $S$ to ${\cal
H}_{c^{\alpha}}$  is stationary.  Stationarity of
$S_{c^{\alpha}}$ imposes no  constraints on the phase
space so that, using some set of coordinates,
 section \ref{uncon} defines an
extension $\{,\}_{c^{\alpha}}$ of the Poisson bracket from
${\cal S}_{c^{\alpha}}$ to ${\cal H}_{c^{\alpha}}$.
To build a Lie bracket on ${\cal H}$ from the $\{,\}_{c^{\alpha}}$,
define
\begin{equation}
\label{gauge broken H bracket}
(F,G)_{\cal H}(p)
\equiv F,_i(p)
\ \{z^i,z^j\}_{c^{\alpha}}(x)
\ G,_j(p)
\end{equation}
for $p=(x,c^{\alpha}) \in {\cal H}$, $x \in \Gamma^I$,
$F$, $G$ any two smooth  functions on
${\cal H}$, and where the functions in the
bracket $\{,\}_{c^{\alpha}}$ are the restrictions of the
indicated functions on ${\cal H}$ to ${\cal
H}_{c^{\alpha}}$.  The Lie bracket properties  of
$(,)_{\cal H}$ follow from those of $\{,\}_{c^{\alpha}}$.
Note that \ref{gauge broken H bracket} depends on the
Lagrange multipliers $\lambda^{\alpha}$  through the
parameterized Poisson bracket $\{,\}_{c^{\alpha}}$ as well
as through $F$ and $G$.

The resulting algebra once again depends
on the choice of coordinates on $\Gamma$ but is invariant
under linear changes of coordinates.  It follows that we
may consistently define $(,)_{\cal H}$ when $\Gamma$ is a
linearized manifold.  Note,
however, that $(,)_{\cal H}$ is completely independent of the
choice of coordinates on $\Lambda^I$.

Because $(A,\phi_{\alpha})_{\cal H}  \neq 0$ for some $A$,
this  algebra
does not have a well-defined
pull back to ${\cal S}$.  It does, however, have a well
defined pull back to
the space ${\cal E} = \cup_{\lambda^{\alpha}} {\cal
S}_{\lambda ^{\alpha}}$ which we will call the space of
canonical evolutions.   This pull back
is independent of smooth coordinate transformations on
either $\Gamma$ or $\Lambda^I$.  In this way, ${\cal E}$
can play a role for constrained systems similar to that of
${\cal S}$ for unconstrained systems and we will use it to
build a quantum theory in \ref{HDQ}. Note that ${\cal E}$
is isomorphic to $\Gamma \times {\Lambda}^I$ as a point in
${\cal E}$ describes the evolution of a point in phase
space under the parameterized Hamiltonian $H +
\lambda^{a}(t) \phi_{a}(t)$.  An equivalent  definition of
${\cal E}$ is therefore ``that subspace of ${\cal H}$  on
which the equations of motion $\{S,_i = 0\}$ are satisfied
but $\{S,_{\alpha}=0\}$ are not."

We now have a Lie bracket $(,)_{\cal H}$ on ${\cal H}$ for
constrained  systems that maps to the Poisson bracket
under pull back to ${\cal E}$ and projection to
${\Gamma}$.  Interestingly, the above procedure defines
such an extension even when some constraints are second
class, though, the presence of second class
constraints would allow us to extend the Dirac bracket
\cite{Dirac} as well.  This option is discussed in
appendix \ref{Dirac}.

\subsection{Gauge Breaking}
\label{gauge fixing}

Following the canonical approach of \cite{Dirac} is not
essential as Lie brackets associated with gauge systems may
also be defined through gauge fixing or other reduced phase
space techniques.  The reduced Poisson algebra ${\cal
A}_H(\Gamma_r)$ extends to ${\cal A}_H({\cal
H}_r)$ on ${\cal H}_r$ by section \ref{uncon}, and we will
see that it extends to ${\cal H}$ as well.  We first
consider the case where ${\cal H}_r$ has been embedded as
a surface ${\cal H}_{gf}$ in ${\cal H}$ by some gauge
fixing procedure.

Much as with our extension of the Poisson bracket from
${\cal S}$ to ${\cal H}$ in section \ref{uncon}, we will
see that this extension depends not so much
on the space ${\cal H}_{gf}$ as on a particular set of
functions (the  gauge fixing functions $P^{\alpha}$) that
define ${\cal H}_{gf}$ through the condition $P^{\alpha} =
c^{\alpha}$ for some $c^{\alpha} \in {\cal R}$.   If these
same functions are used to define a different gauge fixed
slice ${\cal H}_{gf}'$ by $P^{\alpha} = c'^{\alpha}$,
the resulting bracket $(,)_{\cal H}$ on ${\cal H}$ will
not be altered.  However,
if a different set of functions $P'{}^{\alpha}$ are used
to define  the {\it same} surface ${\cal H}_{gf}$,
our construction may define a different bracket
$(,)'_{\cal H}$ on  ${\cal H}$.  It would
thus be incorrect to describe $(,)_{\cal H}$ as a gauge
fixed algebra and  we will refer to it as ``gauge
broken."

We will define
$(,)_{\cal H}$ by introducing a local product structure on
${\cal H}$ that selects gauge fixed {\it local} slices.
Such a structure can be introduced more generally than a
gauge fixed global slice and exists whenever ${\cal H}$ is
a fibre bundle of gauge orbits over ${\cal H}_r$.  When
this  product structure is global, the gauge broken
algebra pulls back to the appropriate gauge fixed algebra
on any cross section.

However, since our general construction will involve local
gauge fixing,  it will be simplest to disentangle locality
from the procedure by first examining the case in which
the product  structure is global.  We will then see that
this analysis could have been performed locally, in
patches, and that the resulting local algebras may be
assembled into a globally defined Lie bracket as
described in Appendix A.

We therefore begin with a system described by an action
$S$ on a space of histories ${\cal H}$ with gauge
invariances indexed at each time by $a \in {\cal G}$.
We assume that this system may be gauge fixed to a slice
${\cal H}_{c^{\alpha}}$ by choosing values $c^{\alpha}$
for a  set of global gauge fixing functions $P^{\alpha}$.
Specifically, we require that the variation of the
restriction  $S_{c^{\alpha}}$
of $S$ to ${\cal H}_{c^{\alpha}}$ has no
gauge invariances and that ${\cal H}_{c^{\alpha}}$ is
transverse to the orbits.  If $S_{c^{\alpha}}$ takes
the canonical form \ref{canact}, it defines a bracket on
${\cal H}_{c^{\alpha}}$ in the manner discussed in section
\ref{uncon} after choosing linearized coordinates
$z^A$ on the gauge fixed phase space $\Gamma_{c^{\alpha}}$.

However, if $S$ is in the canonical form \ref{canact} and
the gauge breaking is canonical (that is, if $P_a(t)$
depends only on $z^A(t)$) then $S_{c^{\alpha}}$
will take the form $\ref{canact}$
only after pull back to a smaller space ${\cal
H}'_{c^{\alpha}}$  in which some of the
equations of motion generated by $S_{c^{\alpha}}$ have
been solved. The relevant equations can be divided into
the constraints $\phi_{\alpha}$ and another set also
indexed by $\Lambda^I$.  This other set arises by varying
$S_{c_{\alpha}}$ with respect to the quantities conjugate
to $P_a(t)$ in the sense of the  canonical Poisson bracket
and which take the form of a further constraint on
$\Gamma_{tc^{a}(t)}\times  \Lambda_t$.  Here,
$\Gamma_{tc^{a}(t)}$ is the subspace of $\Gamma_t$ on
which $P_a(t) = c_a(t)$ and $\Lambda_t$ is the appropriate
copy of $\Lambda$.

These
equations can then be solved for the Lagrange multipliers
$\lambda^a(t)$ and some set of fields $q^a$(t) on
$\Gamma_t$ in the form \begin{mathletters}
\label{can gf}
\begin{equation}
\lambda^{a}(t) = \lambda^a(z^A{}'(t),t)
\end{equation}
\begin{equation}
\label{solve conj}
q^a(t) = q^a(z^A{}'(t),t)
\end{equation}
\end{mathletters}

\noindent
where $z^A{}'$ and $q^a$ are independent functions on
$\Gamma_{tc^{a}(t)}$  so that
the $z^A{}'(t)$ pull back to
coordinates on the phase space $\Gamma'_{tc^{a}(t)}
\subset \Gamma_{tc^{a}(t)}$ in which Eq. \ref{solve conj}
hold.  We assume that $q^{\alpha}$ and $z^i{}'$ on ${\cal
H}_{c^{\alpha}}$ are the pull backs of smooth functions
$q^{\alpha}$ and $z^i{}'$ on ${\cal H}$ and require that
$\{q^a(t),z^A{}'(t)\}_t = 0$ and $\{q^a(t),P^b(t)\}_t =
\gamma^{ab}$ where $\{,\}_t$ is the canonical Poisson
bracket on $\Gamma_t$ and $\gamma^{ab}$ is some matrix
that depends on fields only through $P^{\alpha}$, so that
the choice of  $z^i{}'$ determines $q^{\alpha}$ up to the
choice of $\gamma^{ab}$.   For canonical gauge fixing, the
restriction $S'_{c^{\alpha}}$ to ${\cal H}'_{c^{\alpha}}$
can be written in the form \ref{canact} using the
coordinates $z^A{}'(t)$ so that the methods of section
\ref{uncon} define a Lie bracket $(,)_{c^{\alpha}}$ on
${\cal H}'_{c^{\alpha}}$.  We refer to this type of gauge
breaking as ``based on canonical gauge fixing."

The form of Eq. \ref{can gf} allows $(,)'_{c^{\alpha}}$
to be extended to a bracket $(,)_{c^{\alpha}}$
on ${\cal H}_{c^{\alpha}}$.  To do so, we first introduce
a derivative  operator: $_{;i'}$ that takes derivatives
with respect to $z^i{}'$ along curves of constant
${{\delta S} \over {\delta q^{\alpha}}}\bigg|_{ z^i{}',
P^{\gamma}, \lambda^{\beta}}$ and constant ${{\delta S}
\over {\delta {\lambda}^{\alpha}}}\bigg|_{ z^i{}', P^{\gamma},
q^{\beta}}$.  This poses no difficulties since these two
expressions are ultralocal in time on
${\cal H}_{c^{\alpha}}$.  We then define:
\begin{equation} (F,G)_{c_{\alpha}} (p)
= F_{;i'}(p)\  (z^i{}',z^j{}')'_{c_{\alpha}}
(z'(x)) \
G_{;j} (p)
\end{equation}
for $i' = (A',t)$, $p = (x,c^{\alpha}) \in \Gamma^I \times
\Lambda^I =  {\cal H}_{c^{\alpha}}$, $z'$ the map from
$\Gamma^I$ to $\Gamma'{}^I$ with components $z'^i$, $F$,
$G$ any two smooth functions on ${\cal H}_{c^{\alpha}}$,
and where the functions in the bracket $(,)'_{c^{\alpha}}$
are the restrictions of the indicated functions on ${\cal
H}_{c^{\alpha}}$ to ${\cal H}'_{c^{\alpha}}$.

Whether or not the gauge breaking was canonical,
we now have a Lie bracket
$\{,\}_{c^{\alpha}}$ on each slice ${\cal H}_{c^{\alpha}}$
of a foliation of ${\cal H}$ that was defined using the
pull back to ${\cal H}_{c^{\alpha}}$ of some set of
functions $\phi^{\mu}$ on ${\cal H}$ as coordinates on the
slices.  Together, $\phi^{\mu}$ and $P^{\alpha}$ form
coordinates on ${\cal H}$ so that we can use much the same
technique  as in \ref{constraints} to  define $(,)_{\cal
H}$: \begin{equation}
\label{gb hbracket}
(F,G)_{\cal H}
(p) \equiv F,_{\mu}(p)\
(\phi^{\mu},\phi^{\nu})_{c^{\alpha}} (x)
\ G,_{\nu} (p)
\end{equation}
for $p=(x,c^{\alpha}) \in {\cal H}$, $x \in {\cal
H}_{c^{\alpha}}$, $F$, $G$ any two functions on ${\cal
H}$, and where $F,_{\mu}$ is a derivative with respect to
$\phi^{\mu}$ in $(\phi^{\nu},P^{\alpha})$ coordinates, and
the functions in the bracket $(,)_{c^{\alpha}}$ are the
restrictions to ${\cal H}_{c^{\alpha}}$ of the indicated
functions on ${\cal H}$.  The Lie bracket properties of
$(,)_{\cal H}$ follow from those of $(,)_{c^{\alpha}}$
and  $(,)'_{c^{\alpha}}$.

We now note that this bracket is independent
of linear changes of the coordinates $\phi^{\mu}$ and is
completely invariant under
replacement of the $P^{\alpha}$ by arbitrary functions of
themselves. We may therefore relax our assumption that
$(\phi^{\mu},P^{\alpha})$ form a global coordinate chart
on ${\cal H}$ and that ${\cal H}_{c^{\alpha}}$ is  a
global section transverse to the gauge orbits.  We need
only assume that our $(P^{\alpha},\phi^{\mu})$ coordinates
form a local product structure on ${\cal H}$ -- that is,
they form coordinates in patches on ${\cal H}$ such that
the transition functions map $P's$ to $P's$ and map
$\phi's$ linearly to $\phi's$ and that in each patch
${\cal H}_{c^{\alpha}}$ is a section transverse to the
gauge orbits.

Finally, since every fibre
bundle of gauge orbits over a linearized base space
${\cal H}_r$ of reduced histories has such a
structure, we may use gauge breaking to extend any Lie
algebra on $\Gamma_r$.  We take $\phi^{\mu}$ to be
linearized coordinates  on ${\cal H}_r$
and $P^{\alpha}$ to be functions on the orbits, defined
locally on ${\cal H}$, so that the local slice ${\cal
H}_{c^{\alpha}}$ is  transverse to the orbits.  Since
${\cal H}_r$ is just a  piece of ${\cal H}_r$, ${\cal
A}_H({\cal H}_r)$ defines $(,)_{c^{\alpha}}$  and \ref{gb
hbracket} extends this algebra to ${\cal A}_H({\cal H})$.

\subsection{Properties of Gauge Broken Algebras}
\label{compare}

Note that when the constraints
are first class, the discussion of \ref{constraints} is
identical to the construction of a gauge broken algebra in
\ref{gauge fixing} using the conditions $P^{\alpha} =
\lambda^{\alpha}$ and the global product structure ${\cal
H} = {\cal H}_{\Gamma} \times \Lambda^I$ for the
appropriate space ${\cal H}_{\Gamma} \subset \Gamma^I$.
In this way, both Dirac analysis and gauge fixing may be
regarded as examples of ``gauge breaking."

The canonical case is in fact typical and we can
generalize elements of \ref{constraints} to all gauge
breaking procedures.  For example, we now define a space
${\cal E} = \cup_{c_{\alpha}} {\cal S}_{c_{\alpha}}$ of
gauge broken evolutions where
${\cal S}_{c_{\alpha}}$ is the subspace of ${\cal
H}_{c_{\alpha}}$ on which $S_{c_{\alpha}}$ is stationary.
The extended algebra has a well-defined pull back to
${\cal E}$ which is invariant under any change of
coordinates that respect the local product structure on
${\cal H}$ ({\it i.e.}, that does not mix the $P^{\alpha}$
with the $\phi^{\mu}$).  As before, ${\cal E}$ can be
defined as the subspace on which the equations of motion
$S,_{\mu} = 0$ hold and therefore on which
\ref{can gf} also holds if the gauge breaking is based on canonical
gauge fixing.
As in \ref{constraints}, ${\cal A}_H({\cal H})$ does not
in general have a further pull back to ${\cal S}$.

It may, however, be pulled back to ${\cal S}$ when
the gauge breaking is based on canonical gauge fixing.
To see this when the system is
finite dimensional, recall that since the action is
invariant  under gauge transformations, the equations of
motion are linearly dependent:
\begin{equation}
\label{gauge}
0 = \delta_{\alpha}S = S,_{\beta}
\delta_{\alpha}\lambda^{\beta} + \int_{t_i}^{t_f} dt \
S,_{(A,t)} \delta_{\alpha}z^A(z^B(t)) \end{equation}
where $\delta_{\alpha}$ generates the gauge transformation
labelled by ${\alpha}$,
$S,_{\alpha} = {{\delta S} \over {\delta
\lambda^{\alpha}}}$, $S,_{(A,t)} = {{\delta S} \over
{\delta z^A(t)}}$ and  the notation
$\delta_{\alpha}z^A(z^B(t))$ is to emphasize that the
gauge transformation of $z^A(t)$ depends only on the gauge
parameters and the canonical coordinates $z^B(t)$ at the
same time $t$.   This  equation can be solved to express
${{\delta S} \over {\delta
P^{\alpha}}}\bigg|_{\phi^{\mu}}  = {{\delta z^i} \over
{\delta P^{\alpha}}} \bigg|_{q^{\beta},z'^j} {{\delta S}
\over {\delta z^i}}$ as a function of the  equations of
motion $S,_ {\mu} = {{\delta S} \over {\delta \phi^{\mu}}}
\bigg|_{P^{\alpha}}$ that vanishes when  $S,_{\mu} = 0$.
It follows that in fact all of the equations of
motion in $(\phi^{\mu},P^{\alpha})$ coordinates
vanish on ${\cal E}$.  Thus, ${\cal E}$ and ${\cal S}$
are identical.

Unfortunately, this argument does not go through for
infinite dimensional
systems without considering the details of the boundary
conditions (say, at spatial infinity).  Such conditions
enter when  solving \ref{gauge} for ${{\delta S} \over
{\delta P^{\alpha}}}\bigg|_{\phi^{\mu}}$.
It happens, however, that even in the infinite
dimensional case an algebra defined by canonical gauge
breaking has a well-defined pull back to ${\cal S}$.
A proof of this will be given in
\cite{other} using the methods of generalized Peierls
brackets.

\section{Algebras on spaces of Lagrangian Histories}
\label{Lagrange}

In sections \ref{uncon} and \ref{gauge fixing} we
considered Lie algebras on spaces of
histories associated with the canonical formulation and a phase space
$\Gamma$.  In typical cases,
it is straightforward to
define analogous Lie brackets on any space ${\cal L}$ of
Lagrangian histories, by which we mean the domain space of
an action that may not be in the canonical form
\ref{canact} or \ref{canact gauge}.  This happens when
${\cal L}$ embeds in ${\cal H}$ in the same way as ${\cal
S}$ and ${\cal E}$.

Typical covariant Lagrangians can be derived from the
canonical Lagrangians by
first splitting the coordinates $z^A(t)$ into
configuration variables $q^I(t)$ and momenta $p_I(t)$.
The covariant action is then the restriction of the
canonical action to the subspace ${\cal L}$ on which it is
stationary with respect to variations in the momenta --
effectively solving the equations ${{\delta S} \over
{\delta p_I(t)}} = 0$ to express $p_I(t)$ in terms of the
velocities $\dot{q}^I(t)$ and the Lagrange multipliers
$\lambda^{a}(t)$. Since every algebra on ${\cal H}$ that
we have discussed respects equations of motion that arise
by variation of phase space coordinates, every $(,)_{\cal
H}$ has a well-defined pull back $(,)_{\cal L}$ to ${\cal
L}$.

Note that no constraints have been solved in passing to
${\cal L}$ so that ${\cal L}$ contains both the space
${\cal S}$ of solutions and the spaces ${\cal E}$ of
evolutions defined in \ref{constraints} and \ref{compare}.
As a result, when the Hamiltonian $H$ is quadratic in
momenta, some of the equations of motion that follow from
the restriction $S_{\cal L}$ of $S$ to ${\cal L}$ will be
of less than second order in time derivatives when
expressed in terms of the coordinates $q^I(t)$ and
$\lambda^{a}(t)$ on ${\cal L}$.   The bracket of these
equations of motion with functions on ${\cal L}$ will not
in general vanish, just as was the case on ${\cal H}$.

\section{Quantization}
\label{Quant}

We now investigate quantization of the extended Poisson
algebras defined in \ref{uncon} and \ref{Gauge}.  Our
interest will be focussed on gauge systems and we discuss
gauge-free systems only as a part of the general
introduction.  Following the introductory comments, two
subsections describe quantization of our extended
canonical algebra and of gauge broken algebras, comparing
them with the usual constraint quantization of Dirac
\cite{Dirac} and with gauge fixing methods.

We first note that (at least when global coordinates
exist on ${\cal H}$) any extended Poisson bracket algebra
has a nontrivial center since there are equations of
motion $S,_i = 0$ such that $(S,_i,A) = 0$.  If this
property persists after quantization, the relevant
operators ${\cal O}$ are proportional to the identity:
${\cal O} =  c_{\cal O} \openone$ in any irreducible
representation of the commutator algebra.  The resulting
representation may be thought of as a quantization of the
pull back of $(,)_{\cal H}$ to the subspace on which such
${\cal O}$ take the values $c_{\cal O}$.

We are thus let to consider quantization of the algebra
pulled back to  spaces on which $S,_i = c_i$ for $i$ in
some set ${\cal T} \subset  {\cal I} \times I$ such that
and some $c_i \in {\cal R}$.   However, such spaces
are incompatible with the
dynamics unless $c_i = 0$ so we consider only such spaces
of partial solutions.
Observe that, for the algebras constructed in sections
\ref{uncon}
and \ref{Gauge}, the smallest space of this kind
is either ${\cal E}$ or ${\cal S}$.  For this
reason, we now confine
ourselves to a discussion of $(,)_{\cal E}$ for the
extended canonical bracket and $(,)_{\cal S}$ for gauge
broken algebras based on canonical gauge fixing.

Recall that such a pull back
endows the algebra with additional coordinate invariance
properties.
Also recall that, in general, $(,)_{\cal E}$ and
$(,)_{\cal S}$ have nontrivial centers as well, though
this remaining center can be removed only by pulling back
the algebra further onto a gauge fixed slice.   It will be
convenient not to do so even when such a slice is
available.  In what follows, $I$  is the embedding of
${\cal E}$ or ${\cal S}$ into ${\cal H}$.

\subsection{Canonical Quantization}
\label{HDQ}

We first examine quantization of $(,)_{\cal E}$ defined
in  \ref{constraints} from the canonical Poisson bracket.
Recall that  ${\cal E} = {\Gamma} \times \Lambda^I$
since we have $\dot{z}^A(t) \circ I = \{z^A,H(t) +
\lambda^a(t) \phi_a(t)\}_t \circ I$.  If the constraints
are first class with respect to $(,)_{\cal E}$, we
quantize the algebra  through
\begin{equation}
\label{comms}
[z^A(t),z^B(t)] = i \Omega^{AB}, \
[z^A(t),\lambda^{\alpha}] = 0, \  [\lambda^{\alpha},
\lambda^{\beta}] = 0, \end{equation}
and
\begin{equation}
\label{evolve}
z^A(t_2) = {\cal P} \exp(i \int_{t_1}^{t_2} (H(t) +
\lambda^a(t) \phi_a(t)) dt) \ z^A(t_1) \ {\cal P}\exp(i
\int_{t_2}^{t_1} (H(t) + \lambda^a(t) \phi_a(t)) dt)
\end{equation}  where ${\cal P}$ denotes path ordering.
When some constraints are classically
second class, the Dirac bracket on $\Gamma$
may be used in place of $\Omega^{AB}$
in \ref{comms} so that we quantize the extended Dirac
bracket of appendix \ref{Dirac}.

The constraints are then to be factor ordered in such a
way that they are first class with respect to \ref{comms}
and imposed as conditions that select physical
states as in \cite{Dirac}.  Due to their first class
nature and \ref{evolve}, imposing the entire set
$\{\phi_{\alpha}\}$ of constraints is equivalent to
imposing the constraints $\phi_a(t)$ at any single time
$t$.

Although this strongly
resembles the usual Dirac quantization, the two are not
identical. One difference is that the constraints in
\cite{Dirac} generate arbitrary gauge transformations
but, from \ref{comms}, we have
$[\lambda^{\alpha},\phi_{\beta}] = 0$ even though
$\lambda^{\alpha}$ is not gauge invariant.  Also, as noted
before, any combination $c^{\beta} \phi_{\beta}$ of
constraints can be expressed as a combination of the
constraints at any single time $t$: $c^{\beta} \phi_{\beta}
 = c^a(t) \phi_a(t)$.  Thus,
transformations generated by the constraints may be
parameterized by the values $c^a$ for $a \in {\cal G}$,
while the space of gauge transformations is parameterized
by $c^{\alpha}$ for $\alpha \in  {\cal G}^I$.  In
addition, every transformation generated by the
constraints $\phi^{\alpha}$ extends arbitrarily far to the
future, whereas gauge transformations should have compact
support. More careful consideration shows that the
constraints generate the transformations
$\delta z^i = \epsilon^{\alpha} \delta_{\alpha} z^i$
where $\delta_{\alpha}$ is the gauge transformation
labelled by ${\alpha}$, for those parameters
$\epsilon^{\alpha}$ such that $\epsilon^{\alpha}
\lambda_{\alpha} = 0$.  This is exactly that part of the
gauge freedom not fixed by pulling back to a subspace of
constant $\lambda^{\alpha}$.

Perhaps the most apparent distinction between the above
prescription and that of \cite{Dirac} is that \ref{comms}
considers the Lagrange
multipliers $\lambda^{\alpha}$ to be operators whereas in
\cite{Dirac} they are functions to be specified by hand.
As a result, while any
particular choice of these functions gives a formulation
identical to some irreducible representation of
\ref{comms} and \ref{evolve}, the original prescription of
\cite{Dirac} gives no way to define the evolution of gauge
dependent operators without choosing values for
$\lambda^{\alpha}$ and thereby performing a (partial)
gauge fixing. However, since we have introduced the
operators $z^A(t)$ and $\lambda^a(t)$ for all times $t$,
the evolution of every operator is determined.

This feature is especially useful
in the study of time reparameterization
invariant systems where it allows us to construct gauge
invariant operators by integrating over time.  For
example, in General Relativity, we might be interested in
the total curvature $\int_M d^4x R$ of some four-manifold
$M$ and in the study of the relativistic free particle, we
might be interested in the proper time accumulated between
$x^0 = \alpha$ and $x^0 = \beta$: \begin{equation}
\tau = -\int dt \theta(x^0(t) - \alpha) \theta(\beta-
x^0(t))  \sqrt{-\dot{x}^2(t)}
\end{equation}
or in the value $[x^{\alpha}]_{a_{\mu}x^{\mu} = \tau}$ of
some  coordinate $x^{\alpha}$ when the particle crosses
the hypersurface $a_{\mu} x^{\mu} = \tau$:
\begin{equation}
\label{cross tau}
[x^{\alpha}]_{a_{\mu}x^{\mu} =\tau} = \int dt (a_{\mu}
\dot{x}^{\mu}(t))^{1/4} x^{\alpha}(t)
\delta(a_{\mu}x^{\mu}(t) - \tau) (a_{\mu}
\dot{x}^{\mu})^{1/4} \end{equation}

While such definitions are unlikely to simplify any
computation, they may provide a conceptual advantage over
building operators explicitly through phase space
functions.  For example, we would be free to consider the
commutator algebra defined by \ref{comms} and \ref{evolve}
of all gauge invariants built from the operators $z^i$ and
$\lambda^{\alpha}$.  It is straightforward to show that
this is a Lie algebra, though defined on a monstrously
overcomplete set of  operators and for which explicit
commutation relations are difficult to compute.  However,
these difficulties may now be considered technical
complications to be explored in each model.  As a final
comment we note that a similar construction can be
performed in any  quantization based on an algebra of
functions of histories.

\subsection{Quantization in the presence of a Gribov
Ambiguity} \label{Gribov}

As mentioned in \ref{gauge fixing}, gauge fixed algebras
are given by the pull back of a corresponding gauge broken
algebra when ${\cal H}$ has the structure of a trivial
fibre bundle of gauge orbits.  Thus, gauge fixing and
gauge breaking are nearly equivalent in the absence of a
Gribov ambiguity.  We now investigate the case where  a
Gribov ambiguity {\it is} present.

Recall that such a discussion is possible since a gauge
broken algebra may be defined without reference to global
gauge fixing conditions.  All that is required is for
the $P^{\alpha}$ to form local gauge fixing conditions and that ${\cal H}$
have a linearized structure.
Note, however, that if the $P^{\alpha}$ {\it are} global
gauge fixing conditions then $(A,P^{\alpha}) = 0$ for any
function $A$ on ${\cal H}$.  If the factor ordering
preserves this feature after quantization then
the $P^{\alpha}$ are proportional to the identity operator
in any irreducible representation: $P^{\alpha} =
c^{\alpha} \openone$.   Such representations are
just the irreducible representations of the
algebra pulled back to ${\cal H}_{c^{\alpha}}$ -- i.e.,
the ``gauge fixed representations."

Now, suppose that $P^{\alpha}$ is defined only locally.
In particular, consider a case in which the phase space is
the cotangent bundle over some configuration space ${\cal
Q}$, the gauge conditions $P^{\alpha}$ depend only on the
configuration variables, $P^a(t)$ is independent of time,
and the gauge transformations generate translations on
${\cal Q}$. Similarly, we take the patches on $\Gamma$ to
be cotangent bundles over configuration space patches and
consider corresponding patches on ${\cal H}$ and ${\cal
S}$.  We note that this gauge breaking is based on
canonical gauge fixing so that we may pull back the algebra to ${\cal S}$.

In this case, the bracket $(A,P^{\alpha})_{\cal H}$ is
not defined since $P^{\alpha}$ has not been defined as a
function on ${\cal S}$ but only on some patch $U$.  This
makes quantization of $(,)_{\cal S}$ more difficult.  We
begin by defining:

\begin{equation}
\label{evolve2}
\phi^A(t_2) = {\cal P} \exp(i \int_{t_1}^{t_2} (H(t) +
\lambda^a(t) \phi_a(t)) dt)
\ \phi^A(t_1) \ {\cal P}\exp(i \int_{t_2}^{t_1} (H(t) +
\lambda^a(t) \phi_a(t)) dt)
\end{equation}
and
\begin{equation}
P^a(t) = P^a(t')
\end{equation}
where ${\cal P}$ denotes path ordering.  Thus,
every operator may be built from $\phi^A(t)$ and
$P^{a}(t)$.  Except for $\lambda^a(t)$,
these are functions on $T_*{\cal Q}$, which
are in turn built from
functions and vector fields on ${\cal Q}$.  We then define $\lambda^a(t)$
to be built from functions and vector fields on ${\cal Q}$ through
some factor ordering of \ref{can gf}.

To define the commutator algebra, let $P^{a}_U$ be the
gauge breaking functions on the patch $U$.  Then, for all vector fields
$v_1$ and $v_2$
such that ${\pounds}_{v_1}(P^a_U) = 0 = {\cal
L}_{v_2}(P^a_U)$  on $U$ for all patches $U$, and
all functions $f$ on ${\cal Q}$ we define:
\begin{equation}
\label{comms2}
[v_1,f] = i {\pounds}_{v_1} f \qquad \text{and} \qquad
[v_1,v_2] = i\{v_1, v_2\}_{\pounds}
\end{equation}
where $\{v_1,v_2\}_{\pounds}$ is the Lie bracket of
vector fields. Additionally, if $q_{Ua} =
(\gamma^{-1})_{ab} q^a_U$ is the function on $U$ conjugate
to $P^a_U$ in the sense of the canonical Poisson bracket,
then for any function $\rho_U$ on ${\cal Q}$ with support
on $U$ we define: \begin{equation} [\rho_{U} q_{aU},A] =
(\gamma^{-1})_{ab} [\rho_U q^b(z^A{}'),A] \end{equation}
where $A$ is some function on $T_*{\cal Q}$ and
$q^b(z^A{}')$ is some factor ordering of the solution in
\ref{can gf}.  Note that the entire algebra is defined
through the action of vector fields on functions by
infinitesimal translation.  However, because the algebra
is degenerate, not  all infinitesimal translations are
generated.

Thus, ${\cal Q}$ may be partitioned into equivalence
classes such that points in each class
are joined by a series of infinitesimal translations
generated by \ref{comms2}.  The set of functions with
support on any  such equivalence
class carries a representation of our algebra.  If, in
addition, there is an open set ${\cal V} \neq {\cal Q}$
in ${\cal Q}$ such that every point $p \in {\cal V}$
lies in some equivalence class ${\cal C}_p$ that is
entirely  contained in ${\cal V}$,
then any
representation carried by $L^2$ functions on ${\cal
Q}$ in which the vector fields act by translations
generated by \ref{comms2} and the functions act by
multiplication is topologically reducible as well.
Regardless of this, if ${\cal Q}$ is a fibre bundle of
gauge orbits over some base space ${\cal B}$, gauge
transformations act on  functions with support a single
equivalence class only through a representation of
$\pi_1({\cal B})$.  Thus, the action of the gauge group
may be much simpler on an irreducible representation than
on the full configuration space ${\cal Q}$

\section{Discussion}

We have seen that the Poisson bracket can be extended to a
Lie bracket of functions on the space ${\cal H}$ of
canonical histories and on spaces ${\cal L}$ of Lagrangian
histories.  For gauge  systems, we extended both the
canonical Poisson bracket and reduced phase space Poisson
bracket, observing that both are examples of ``gauge
breaking."  Gauge breaking is an interesting technique in
itself, as it resembles gauge fixing yet may be performed
in the presence of a Gribov ambiguity.

We then investigated quantization of such algebras.  This
reduces to a study of $(,)_{\cal E}$ and $(,)_{\cal S}$.
We found that while gauge breaking may produce constraints
that resemble those of \cite{Dirac}, the interpretation
of these constraints is different and they generate
residual symmetry transformations instead of gauge
transformations.  We also saw that the Heisenberg picture
nature of our quantized algebra allows the construction of
invariant operators in time reparameterization invariant
systems by integration over time and that a quantum theory
based on a gauge broken algebra in the presence of a
Gribov ambiguity may  still be reducible to a
representation in which the gauge symmetry acts simply.

All of this was intended, however, to set the stage for
\cite{other}. We have presented an introductions to
algebras on ${\cal H}$ and their pull backs while
introducing the space ${\cal E}$ of evolutions and the
concepts of locally defined algebras and gauge broken
algebras. Since our construction was based on the Poisson
bracket, we were also able to provide a straightforward
comparison with more familiar techniques.  In
\cite{other}, we will place these ideas in the more
general and unified framework of the generalized Peierls
algebra.

\acknowledgements
This work was partially supported by a National Science
Foundation Graduate Fellowship, by NSF grants
PHY90-05790 and PHY93-96246, and by research funds provided by Syracuse
University.  The author would like to express his thanks
to Carlos Ordo\~nez, Josep Pons, and the Syracuse
relativity group for their encouragement and to his
thesis advisor Bryce DeWitt.

\appendix
\section{Locally Defined Algebras}
\label{assemble}
In this appendix we define a Lie bracket of functions
on a manifold $M$
given a set of Lie brackets $\{,\}_i$ of functions $F_i$
whose support lies in patches $U_i$ when these
algebras are compatible in the overlap regions and the
patches cover $M$.  Our specific
compatibility assumption is that for two
such brackets $\{,\}_i$ and $\{,\}_j$ on patches $U_i$ and
$U_j$ and all smooth functions $F$ and $G$ with
support in $U_i \cap U_j$, we have
$\{F,G\}_i = \{F,G\}_j$.  Note
that $supp(\{F,G\}_i) \subset supp(F) \cap supp(G)$.

Now, given any functions $F$ and $G$ on $M$, we write
\begin{equation}
\label{decomp}
F = \sum_{patches} F_i, \qquad \qquad
G= \sum_{patches} G_i
\end{equation}
where the supports of $F_i$ and $G_i$ both lie in the
patch $U_i$. We then define the bracket of $F$ and $G$ by
\begin{equation}
\label{global}
(F,G) \equiv \sum_{patches} \{F_i,G_i\}_i
\end{equation}
Note that our compatibility
assumption guarantees that \ref{global} is
independent of the decomposition \ref{decomp}.

\section{Extension of the Dirac Bracket}
\label{Dirac}

In this appendix, we describe how the techniques of
\ref{uncon} may also be used to extend the Dirac bracket
\cite{Dirac} to ${\cal H}$ when the constraints are
entirely second class.  We consider an action of the form
\begin{equation}
S = \int_{t_1}^{t_2} [\case{1}{2}
\Omega^{-1}_{AB}z^A\dot{z}^B -H -\lambda^a \xi_a]
\end{equation}
for $\Omega^{AB}$, $z^A$, and $H$ as in \ref{uncon} and
some $\xi_a(t) = \xi_a(z^A(t),t)$.  The $\lambda^a(t)$ are
Lagrange multipliers that enforce the constraints
$\xi_a(t) = 0$.  If the constraints are second class, the
matrix $\Delta_{ab} = \xi_{a|A} \Omega^{AB} \xi_{b|B}$ is
invertible.  The Dirac bracket $\{,\}^D$  is then defined
\cite{Dirac} by \begin{equation}
\{z^A,z^B\}^D = \Sigma^{AB} = \Omega^{AB} - \Omega^{AC}
\xi_{a|C} (\Delta^{-1})^{ab} \xi_{b|D} \Omega^{DB}
\end{equation}

While the Dirac bracket is not defined on $\lambda^a$, the
equations of motion $\xi_a(t) = 0$ place a constraint
$\xi_{a|A}(t)(z^A(t),z^B(t'))^D_{\cal H}=0$ on any
extension $(,)^D_{\cal H}$ of $\{,\}^D$.  These two
features combine in such a way that the requirements
$(S,_i,A)^D_{\cal H} = 0$ and $(z^A(t),z^B(t))^D_{\cal H}
= \Sigma^{AB}$ uniquely define the extension $(,)^D_{\cal
H}$ to be given by:
\begin{mathletters}
\begin{equation}
(z^A(t_1),z^B(t_2))^D_{\cal H} = T^D_L(t_2,t_1)^A_C
\Sigma^{CB}
\end{equation}
\begin{eqnarray}
(\lambda^a(t_1),z^B(t_2))^D_{\cal H} &=& - \chi^{aA}(t_1)
[\Omega^{-1}_{AC} {{\partial} \over {\partial t_1}}
(z^C(t_1),z^B(t_2))^D_{\cal H} \cr &+&  (H_{|AC}(t_1) +
\lambda^a(t_1)\xi_{a|AC}(t_1))
(z^C(t_1),z^B(t_2))^D_{\cal H}]
\end{eqnarray}
\begin{eqnarray}
(\lambda^a(t_1),z^B(t_2))^D_{\cal H} &=& - \chi^{aA}(t_1)
[\Omega^{-1}_{AC} {{\partial} \over {\partial t_1}}
(z^C(t_1),\lambda^b(t_2))^D_{\cal H} \cr &+& (H_{|AC}(t_1) +
\lambda^a(t_1)\xi_{a|AC}(t_1))
(z^C(t_1),\lambda^b(t_2))^D_{\cal H}]
\end{eqnarray}
where
\begin{equation}
T_L^D(t_2,t_1)^A_C = {\cal P} \exp[\int^{t_1}_{t_2}dt \
{\bf Q}(t)]^A_C \ ,
\end{equation}
\begin{equation}
{\bf Q}^A_C = \Sigma^{AB} (H_{|BC} + \lambda^a \xi_{a|BC})
- \Omega^{AB} \xi_{a|B} \Delta^{-1ab} {{\partial} \over
{\partial t}} \xi_{b|C}
\end{equation}
and
\begin{equation}
\chi^{bA} \xi_{a|A} = \delta^b_a \ .
\end{equation}
We note that such a $\chi^{aA}$ exists when the
constraints $\xi_a$ are independent.
\end{mathletters}

As with the
Poisson bracket, the extended Dirac bracket may be pulled
back to spaces of partial solutions.  In particular, such
pull backs may be used to define the Dirac bracket on
spaces ${\cal L}$ of  Lagrangian histories.  Similarly,
when both first and second class constraints are present,
the techniques of section \ref{Gauge} may be used to
define extensions either of the canonical or reduced Dirac
bracket.


\begin{references}

\bibitem[*]{Marolf} Electronic address:
marolf@hbar.phys.psu.edu.

\bibitem{other}  D. Marolf {\it preprint} The Pennsylvania State
University, (1993), CGPG-93/8-5, hep-th/9308150, submitted to
Ann. Phys. {\it N.Y.}.

\bibitem{Peierls} R. E. Peierls {\it Proc. Roy. Soc. (London)}
{\bf 214} (1952), 143-157.

\bibitem{Dirac}  P. A. M. Dirac  ``Lectures on Quantum
Mechanics," Belfor Graduate School of Science, Yeshiva
University, New York, 1964.

\bibitem{Bryce}  B. DeWitt {\it in} ``Relativity, Groups,
and Topology II: Les Houches 1983"
(B. DeWitt and R. Stora, Ed.),  part2, p. 381,
North-Holland, New York, 1984.

\bibitem{Wald}  J. Lee and R. M. Wald {\it J. Math.
Phys.} {\bf 31} (1990), 725-743.

\bibitem{Abhay} A. Ashtekar {\it in} ``Mechanics,
Analysis, and Geometry: 200 years after Lagrange" (M.
Francaviglia, Ed.), Elsevier Science Publishing, New York, 1991.

\bibitem{Witten}  \v{C}. Crnkovi\'c and E. Witten {\it in}
``Newton's tercentenary volume" (S. W. Hawking and
W. Israel, Ed.), Cambridge Univ. Press, Cambridge, 1987.


\end{references}
\end{document}